# Electronic structure of rare-earth infinite-layer ReNiO$_2$ (Re=La, Nd)

Peiheng Jiang[1,†], Liang Si[1,2,†], Zhaoliang Liao[3], Zhicheng Zhong[1,4*]

[1] *Key Laboratory of Magnetic Materials and Devices & Zhejiang Province Key Laboratory of Magnetic Materials and Application Technology, Ningbo Institute of Materials Technology and Engineering, Chinese Academy of Sciences, Ningbo 315201, P. R. China*

[2] *Institut für Festkörperphysik, TU Wien, Wiedner Hauptstraße 8-10, 1040 Vienna, Austria*

[3] *National Synchrotron Radiation Laboratory, University of Science and Technology of China, Hefei 230026, P. R. China*

[4] *China Center of Materials Science and Optoelectronics Engineering, University of Chinese Academy of Sciences, Beijing 100049, P. R. China*

[*]*Corresponding authors: Z.Z. (email: zhong@nimte.ac.cn)*

[†] *These authors contributed equally to this work.*

**Abstract:** The discovery of infinite layer nickelate superconductor marks the new era in the field of superconductivity. In the rare-earth (Re) nickelates ReNiO$_2$, although the Ni is also of $d^9$ electronic configuration, analogous to Cu $d^9$ in cuprates, whether electronic structures in infinite-layer nickelate are the same as cuprate and possess the single band feature as well are still open questions. To illustrate the electronic structure of rare-earth infinite-layer nickelate, we perform first principle calculations of LaNiO$_2$ and NdNiO$_2$ compounds and compare them with that of CaCuO$_2$ using hybrid functional method together with Wannier projection and group symmetry analysis. Our results indicate that the Ni-$d_{x^2-y^2}$ in the LaNiO$_2$ has weak hybridization with other orbitals and exhibits characteristic single band feature, whereas in NdNiO$_2$, the Nd-$f$ orbital hybridizes with Ni-$d_{x^2-y^2}$ and is a non-negligible ingredient for transport and even high-temperature superconductivity. Given that the Cu-$d_{x^2-y^2}$ in cuprate strongly hybridizes with O-2$p$, the calculated band structures of nickelate imply some new band characters which is worth to gain more attentions.

**Main text:**

Since the discovery of high-$T_C$ cuprate superconductor in 1986 [1], searching for new high-$T_C$ superconductor and revealing the mechanism of high-$T_C$ superconductivity (HTSC) are invariably at the forefront of research of condensed matter physics. Although the mechanism of HTSC in the cuprates remains a puzzle, there is a general consensus that the key structural unit $CuO_2$ plane, which exhibits characteristic quasi-two-dimensional (2D) physics of the $d_{x^2-y^2}$ orbital, plays a central role in HSTC. The Cu in cuprate has $d^9$ electron configuration and the crystal field of $CuO_2$ plane strongly lifts the degeneracy of the $e_g$ orbital with $d_{x^2-y^2}$ located above $d_{z^2}$. The $d_{x^2-y^2}$ hybridizes with oxygen $2p$ bands, forming an effective singlet band with zero spin moment known as the Zhang-Rice singlet (ZRS) [2]. The hole in oxygen site couples antiferromagnetically with hole on the copper site. The hole doping can suppress the antiferromagnetic ordering and drive the cuprate into superconductive phase wherein the Cooper pairs are proposed to be mediated by antiferromagnetic spin fluctuation [3,4].

Designing and fabricating an analogue of cuprate superconductor is one of the key steps towards identifying the key ingredients and unraveling the mechanism of HSTC. Nickelates are one of the ideal systems to mimic cuprates. In bulk perovskite $ReNiO_3$ system, the Ni is of $d^7$ configuration with one electron in the doubly degenerate $e_g$ orbitals. One idea to generate cuprate-like electronic structure is to remove the degeneracy of $e_g$ orbitals and deplete the $d_{z^2}$ orbital, which leads to a half-filled $d_{x^2-y^2}$ band similar to the parent compound of cuprates. This is the idea behind the theoretical proposal of $LaAlO_3/LaNiO_3$ hetero-structures by Chaloupka *et al.* in 2008 [5]. The actual orbital polarization obtained by experimental efforts, however, is too small to produce characteristic $d_{x^2-y^2}$ single-band electronic structure [6,7].

To remove the apical oxygen in nickelates and then flip the orbital configuration to lift up the $d_{x^2-y^2}$ against the $d_{z^2}$ is another path to mimic orbital configuration in cuprates [8-14]. In this case, to hall-fill the highest-energy $d_{x^2-y^2}$ orbital, Ni needs to be $d^9$. One promising candidate is the infinite layer $ReNiO_2$, which can be obtained by reducing the ReNiO$_3$. In infinite layer ReNiO$_2$, the Ni$^{1+}$ possesses $d^9$ configuration with half-filled $d_{x^2-y^2}$. Exploring the superconductivity in ReNiO$_2$ starts earlier from 2009 in (La,Sr)NiO$_2$ [15], but no evidence of superconductivity in (La,Sr)NiO$_2$ is found experimentally. A breakthrough is made by Li *et al.* who discovered the superconductivity in Nd$_{0.8}$Sr$_{0.2}$NiO$_2$ with $T_C$ ranging from 9-15 K depending on the samples [14]. The discovery of superconducting Nd$_{0.8}$Sr$_{0.2}$NiO$_2$ sheds the light on the era of nickelate superconductivity, and is now motivating worldwide efforts to explore this new system [16-26]. A systematic comparison between infinite layer nickelate with cuprate can pave the way to identify essential features for superconductivity and to narrow down the range of possible theory of HSTC. The first glance of the nickelate in comparison with the cuprate is what the difference of their electronic structures is. Whether the Zhang-Rice singlet band model is still valid for nickelate? Additionally, to what extend do the *d* and *f* orbital of the rare earth—which are close to the Fermi level—affect the electronic structure and participate in the superconductivity?

With these questions in mind, we perform first principle density functional theory (DFT) calculations using hybrid functionals which incorporate 25% short range Hatree-Fock exchange, 75% full-range Perdew–Burke–Ernzerhof (PBE) exchange, and 100% PBE correlation [27]. The hybrid functional performs exceptionally well in predicting band-gap of semiconductors such as Si and SrTiO$_3$, and describing systems with a mixture of itinerant and localized electrons from different orbitals, where the delicate balance between itinerancy and localization can produce multiple competing ground states and phases [28]. For the rare-earth oxides of LaNiO$_2$ and NdNiO$_2$, both the transition metal *d* electrons and lanthanide *f* electrons participate in electronic band structure near Fermi level [11,12]. The hybrid functional which can deal with both itinerant and localized electrons without any adjustable parameters is a very effective method to describe the electron behavior of such materials. By performing the first principle DFT calculation and group symmetry analysis, we show that the Ni-$d_{x^2-y^2}$ in the LaNiO$_2$ has very weak hybridization with other orbitals and exhibits characteristic single band feature, whereas in NdNiO$_2$ the Nd-$f$ orbital hybridizes with Ni-$d_{x^2-y^2}$ and makes the NdNiO$_2$ different from both LaNiO$_2$ and cuprates. These facts imply that a single-band model may not capture all the key ingredients of NdNiO$_2$.

We performed the DFT calculations within the generalized gradient approximation [29] and the projector augmented wave method as implemented in the Vienna ab-initio simulation package (VASP) [30,31]. To confirm the reliability of our results, the main calculations have been checked by Wien2K code [32] with plane wave method and full potential. For La and Nd, 10 and 14 valence electrons of $5s^26p^65d6s^2$ and $5s^26p^64f^45d6s^2$ were considered in calculations, respectively. We performed spin-polarized calculations for NdNiO$_2$, while complex spin order is not considered for simplicity. The PBEsol functional was adopted for structure relaxation. The relaxed lattice constants are $a = b = 3.88$ Å and $c = 3.35$ Å for LaNiO$_2$, $a = b = 3.86$ Å and $c = 3.24$ Å for NdNiO$_2$. A hybrid density functional of Heyd–Scuseria–Ernzerhof (HSE06) [27], which includes the short range exact exchange function and thus can deal with both itinerant and localized electrons, was used for an accurate prediction of electronic band structure. The Brillouin zone was sampled with 12×12×12 and 8×8×8 k-mesh for DFT and HSE calculations, respectively. The kinetic energy cutoff was set to 500 eV. To obtain the tight-binding model Hamiltonian, the La-5$d$ and Ni-3$d$ DFT bands are projected onto maximally localized Wannier functions using WIEN2WANNIER [33] and VASP2WANNIER interfaces.

The infinite layer ReNiO$_2$ is isostructural to CaCuO$_2$, which can be derived from removing the apical oxygen of the octahedra in ReNiO$_3$. In detail, it is stacked by Re layers and NiO$_2$ planes alternately in a tetragonal lattice. The Re atom is surrounded by an oxygen tetragon, and the Ni atom is surrounded by a Re tetragon and also by oxygen square in the NiO$_2$ plane. The local point group is $D_{4h}$, $D_{4h}$ and $D_{2h}$ for La (Nd), Ni and O, respectively. With this lattice structure, we first investigate the orbital projected electronic band structure of LaNiO$_2$ where the La-$f$ orbital can be neglected and only La-$d$ orbital has to be considered. As shown in **Fig. 1(a)**, six bands within the energy range of −10 ~ −5 eV are contributed by O-$p$ orbitals, and four bands of Ni-$t_{2g}$ and $d_{z^2}$ orbitals are located at a higher range of −5 ~ −2 eV. Near the Fermi level between −2 ~ 3 eV, the bands are mainly contributed by Ni-$d_{x^2-y^2}$, La-$d_{xy}$ and La-$d_{z^2}$ orbitals. Other La-$d$ orbitals and all La-$f$ orbitals form the upper bands.

The prominent band and orbital characters of LaNiO$_2$ can be further characterized by the Fermi surface and partial charge density. As shown in **Fig. 2(a)**, there are two Fermi spheres around Γ and A points with 0.10 and 0.19 electrons included, respectively, and one large electron pocket around Γ-Z direction, which are contributed by La-$d_{z^2}$, La-$d_{xy}$, and Ni-$d_{x^2-y^2}$ orbitals, respectively. In particular, the cylindrical distribution of Fermi surface and energy dispersion of Ni-$d_{x^2-y^2}$ orbital indicate a quasi-2D character along Γ-Z direction. **Figures 2(b)-(d)** display the charge density of the pockets around Γ, Z and A points, respectively. These localized charge distribution around Z point and large orbital-weight of the Ni-$d_{x^2-y^2}$ band together indicate the localization of this orbital. For the other pocket around A point, the charge disperses to almost the whole space of the La atomic plane, which results from the itinerant La-$d_{xy}$ orbital. Such a non-localized character is consistent with the smaller orbital-weight as shown in **Fig. 1(a)**.

The orbital hybridizations are essential properties to estimate electronic structures and to investigate the single-band picture. Although the Ni-$d_{x^2-y^2}$, La-$d_{xy}$ and La-$d_{z^2}$ orbitals are all located near the Fermi level, the weak hybridization between Ni-$d_{x^2-y^2}$ and other orbital as revealed by our Wannier projection tight binding calculation and group symmetry analysis demonstrate that the LaNiO$_2$ can be well described by single band picture as we discussed in the following. The hybridization between two orbitals is mainly determined by two parameters. One is the energy splitting of two orbitals Δ. The other parameter is the hopping term $t$, which is related to orbital shape and lattice symmetry. The strength of hybridization is expressed by the formula $\sqrt{1+(\frac{2t}{\Delta})^2}-1$. A small Δ and a large $t$ will lead to strong hybridization. In the case of weak hybridization, the system can be simplified as a single band. When the hybridization becomes strong, extra treatment of the band and theoretic model should be considered. A typical example is the modified effective single band model, such as Zhang-Rice singlet [2] which mixes the effective information of $p$ orbital into $d$ orbital. Another example is Emery model [34,35], which is a three-band Hubbard model with one Cu-$d_{x^2-y^2}$ and two O-$p$ orbitals.

With respect to LaNiO$_2$, our calculations show that the Ni-$d_{x^2-y^2}$ has negligible hybridization with O-$p$ orbitals, which is in strong contrast to the case in cuprate. Using the CaCuO$_2$ as an example, the Cu-$d_{x^2-y^2}$ orbital has strong hybridization with O-$p$ orbital owing to small $\Delta$ and a large hopping term $t$. In this situation, the system is generally described by an effective $d$-$p$ hybrid single band model. While in LaNiO$_2$, the energy splitting $\Delta$ between Ni-$d_{x^2-y^2}$ and O-$p$ orbitals is as big as 6 eV. Meanwhile, the orbital weight of $d_{x^2-y^2}$ in LaNiO$_2$ is much larger than CaCuO$_2$, e.g. 36% larger at A point. Both of these two facts reveal a much weaker orbital hybridization in LaNiO$_2$ than that in CaCuO$_2$. As a consequence, the $d$-$p$ singlet model is not necessary for LaNiO$_2$. What is more, the Ni-$d_{x^2-y^2}$ orbital is also found to not hybrid with other Ni-$d$ orbitals. First, the Ni-$d_{x^2-y^2}$ and Ni-$t_{2g}$ hybridization is prohibited by orbital orthogonality as revealed by the zero hopping term between Ni-$d_{x^2-y^2}$ and Ni-$t_{2g}$ (see Supplemental Material, SM [36]). Secondly, the inter-orbital hybridization between Ni-$d_{x^2-y^2}$ and the Ni-$d_{z^2}$ is energetically unfavorable due to large energy splitting $\Delta$ in spite of nonzero hopping term $t$ of 0.035 eV between Ni-$d_{x^2-y^2}$ and Ni-$d_{z^2}$.

Our results also indicate weak hybridization between Ni-$d_{x^2-y^2}$ and La-$d$ orbital. The hopping terms from Ni-$d_{x^2-y^2}$ to La-$d_{xy/z^2}$ orbitals are zero due to the orbital orthogonality [36]. Moreover, when wave vector approaches $k_x = 0.5$ ($\Gamma$-X path), the wave vector group is reduced from $D_{4h}$ ($\Gamma$) to $C_{2v}$ ($\Gamma$-X) and the Ni-$d_{x^2-y^2}$ band is described by $A_1$, which is the same with La-$d_{z^2}$ band. The hybridization between Ni-$d_{x^2-y^2}$ and La-$d_{z^2}$ is allowed in terms of their same symmetry. Yet, a large $\Delta$ between Ni-$d_{x^2-y^2}$ and La-$d_{z^2}$ severely reduces hybridization. Along X-M-$\Gamma$ path, the Ni-$d_{x^2-y^2}$ band is described by $B_1$ ($C_{2v}$) while the La-$d_z^2$ band is described by $A_1$ ($C_{2v}$), the hybridization is prohibited by the difference in symmetry. When these two orbitals approach A point ($k_x = k_y = k_z = 0.5$), the characters and symmetric properties of the bands do not change: the Ni-$d_{x^2-y^2}$ band hosts representation of $A_2$ ($C_{2v}$) and the La-$d_{xy}$ band is described by $B_1$ ($C_{2v}$). The most striking consequence of the weak/no hybridization of Ni-$d_{x^2-y^2}$ with other orbitals is that no hybridization gap (band anti-crossing) is formed in the whole Brillouin zone near Fermi level. Further analysis indicates that at $k_z = 0.5$ plane a nodal line is formed and protected by crystal symmetry. The corresponding symmetry analysis can be found in SM [36]. With quasi-single-band behavior of Ni-$d_{x^2-y^2}$, tight binding Hamiltonian is constructed through Wannier projection [36].

By replacing La with Nd, the partially filled *f* electrons are introduced into the system. **Fig. 3** shows the orbital projected electronic band structure of NdNiO$_2$. Here the spin polarization is considered because of the large local magnetic moment of Nd-*f* electrons. The total magnetic moment is 3.91 $\mu_B$ per unit cell, and the local moments of Nd-*f* and Ni-*d* are 3.03 $\mu_B$ and 0.89 $\mu_B$, respectively. As shown in **Fig. 3**, there is a large splitting between spin majority and minority channels of Ni-$d_{x^2-y^2}$ orbital. The Nd-*d* orbitals still cross Fermi level and self-dope the system like La-*d* orbitals in LaNiO$_2$. We now focus on the spin majority channel and find that the Nd-*f* orbitals are much more localized and almost do not disperse. Three *f* bands below Fermi level and one *f* band cross Fermi level contribute a local moment of 3.03 $\mu_B$. Since *f* electrons are located near Fermi level, a *d-f* hybridization occurs at M point. This hybridization is so strong that the *d*-orbital is split and a gap is opened. In contrast, such hybridization does not occur at A point. Unlike the quasi-single-band behavior in LaNiO$_2$ and CaCuO$_2$, the band dispersion of NdNiO$_2$ near Fermi energy exhibits an unusual behavior: the dominating Ni-$d_{x^2-y^2}$ band shows strong hybridization with *f* orbitals around M point, whereas it exhibits quasi-single-band behavior around A point. These characters make the electronic structures of NdNiO$_2$ differ strongly from LaNiO$_2$. The physical origin of this unusual phenomenon is explained by our symmetry analysis, for details please see Fig. S6 [36]. We can draw a conclusion that there is strong hybridization between *f* and *d* electrons in NdNiO$_2$.

In summary, we have performed the DFT calculations of infinite layer ReNiO$_2$ (Re=La, Nd) with hybrid functional method and group symmetry analysis. Our results suggest that the Ni-$d_{x^2-y^2}$ sub-band in LaNiO$_2$ can be treated as an isolated single band due to its weak/no hybridization with other orbitals. As a consequence, the 2D single band character in LaNiO$_2$ becomes even stronger than that of CaCuO$_2$. Meanwhile, the La-$d$ electrons have two non-negligible impacts. One is the self-doping effect, which might lead to the redistribution of charge and alter the magnetic properties of the system. The second effect is that the interaction between La-$d$ electrons and Ni-$d$ electrons might be substantial. The itinerant La-$d$ orbitals provide an electron background in LaNiO$_2$, while the 2D single band of Ni-$d_{x^2-y^2}$ orbital is localized with the energy near Fermi level, and only weak hybridization exists between them.

For NdNiO$_2$, the Nd-$d$ orbital behaves similarly with La-$d$ orbital in LaNiO$_2$. Nd-$d$ orbital also crosses Fermi level and has self-doping effect. The hybridization between Nd-$d$ orbital and Ni-$d_{x^2-y^2}$ is also very weak. The big difference between NdNiO$_2$ and LaNiO$_2$ arises from their $f$-orbital. Different from the LaNiO$_2$ wherein the $f$ orbitals are far from Fermi level and Ni-$d$ orbitals, the Nd-$f$ orbitals are located near Fermi level and strongly hybridize with Ni-$d_{x^2-y^2}$ in NdNiO$_2$. Additionally, the large magnetic moment of Nd-$f$ orbitals can lead to spin polarization of Ni orbital and would affect the magnetic ordering of NdNiO$_2$. The magnetization of $f$ electrons would also make the single Ni-$d_{x^2-y^2}$ model not as effective as that in LaNiO$_2$. The role of rear-earth $f$ orbital in affecting the electronic structure then suggests a new twist of freedom to engineer superconductive and magnetic behavior in infinite layer nickelates by engineering the $f$-$d$ orbital hybridization, which is usually not feasible in cuprate.

**Acknowledgements:**

We gratefully acknowledge financial support from the National Key R&D Program of China (2017YFA0303602), 3315 Program of Ningbo, and the National Nature Science Foundation of China (11774360, 11904373). P.J. was also supported by the China Postdoctoral Science Foundation (Grant No. 2018M642497). L.S. was also supported by the European Research Council (ERC) under the European Union's

Seventh Framework Program (FP/2007-2013) through ERC Grant No. 306447, and by the Austrian Science Fund (FWF) through project P 30997. Research by Z.L. was funded by Hundred Talent Program of the Chinese Academy of Sciences and supported by the Fundamental Research Funds for the Central Universities. Calculations were performed at the Supercomputing Center of Ningbo Institute of Materials Technology and Engineering.

**Reference:**

[1] J. G. Bednorz and K. A. Muller, Z. Phys. B **64**, 189 (1986).

[2] F. C. Zhang and T. M. Rice, Phys. Rev. B **37**, 3759 (1988).

[3] P. W. Anderson, Phys. Rev. Lett. **96**, 017001 (2006).

[4] P. W. Andersen, Science **235**, 1196 (1987).

[5] J. Chaloupka and G. Khaliullin, Phys. Rev. Lett. **100**, 016404 (2008).

[6] E. Detemple, Q. M. Ramasse, W. Sigle, G. Cristiani, H. U. Habermeier, E. Benckiser, A. V. Boris, A. Frano, P. Wochner, M. Wu, B. Keimer, and P. A. van Aken, Appl. Phys. Lett. **99**, 211903 (2011).

[7] P. Hansmann, X. Yang, A. Toschi, G. Khaliullin, O. K. Andersen, and K. Held, Phys. Rev. Lett. **103**, 016401 (2009).

[8] M. Hepting, R. J. Green, Z. Zhong, M. Bluschke, Y. E. Suyolcu, S. Macke, A. Frano, S. Catalano, M. Gibert, R. Sutarto, F. He, G. Cristiani, G. Logvenov, Y. Wang, P. A. van Aken, P. Hansmann, M. Le Tacon, J. M. Triscone, G. A. Sawatzky, B. Keimer, and E. Benckiser, Nat. Phys. **14**, 1097 (2018).

[9] Z. Liao, E. Skoropata, J. W. Freeland, E. J. Guo, R. Desautels, X. Gao, C. Sohn, A. Rastogi, T. Z. Ward, T. Zou, T. Charlton, M. R. Fitzsimmons, and H. N. Lee, Nat. Commun. **10**, 589 (2019).

[10] M. A. Hayward, M. A. Green, M. J. Rosseinsky, and J. Sloan, J. Am. Chem. Soc. **121**, 8843 (1999).

[11] K. W. Lee and W. E. Pickett, Phys. Rev. B **70** (2004).

[12] V. I. Anisimov, D. Bukhvalov, and T. M. Rice, Phys. Rev. B **59**, 7901 (1999).

[13] A. Ikeda, T. Manabe, and M. Naito, Physica C: Superconductivity **495**, 134 (2013).

[14] D. Li, K. Lee, B. Y. Wang, M. Osada, S. Crossley, H. R. Lee, Y. Cui, Y. Hikita,

and H. Y. Hwang, Nature **572**, 624 (2019).

[15] M. Kawai, S. Inoue, M. Mizumaki, N. Kawamura, N. Ichikawa, and Y. Shimakawa, Appl. Phys. Lett. **94**, 082102 (2009).

[16] A. S. Botana and M. R. Norman, arXiv:1908.10946 (2019).

[17] H. Sakakibara, H. Usui, K. Suzuki, T. Kotani, H. Aoki, and K. Kuroki, arXiv:1909.00060 (2019).

[18] J. E. Hirscha and F. Marsiglio, arXiv:1909.00509 (2019).

[19] M. Jiang, M. Berciu, and G. A. Sawatzky, arXiv:1909.02557 (2019).

[20] M. Hepting, D. Li, C. J. Jia, H. Lu, E. Paris, Y. Tseng, X. Feng, M. Osada, E. Been, Y. Hikita, Y.-D. Chuang, Z. Hussain, K. J. Zhou, A. Nag, M. Garcia-Fernandez, M. Rossi, H. Y. Huang, D. J. Huang, Z. X. Shen, T. Schmitt, H. Y. Hwang, B. Moritz, J. Zaanen, T. P. Devereaux, and W. S. Lee, arXiv:1909.02678 (2019).

[21] X. Wu, D. D. Sante, T. Schwemmer, W. Hanke, H. Y. Hwang, S. Raghu, and R. Thomale, arXiv:1909.03015 (2019).

[22] Y. Nomura, M. Hirayama, T. Tadano, Y. Yoshimoto, K. Nakamura, and R. Arita, arXiv:1909.03942 (2019).

[23] J. Gao, Z. Wang, C. Fang, and H. Weng, arXiv:1909.04657 (2019).

[24] S. Ryee, H. Yoon, T. J. Kim, M. Y. Jeong, and M. J. Han, arXiv:1909.05824 (2019).

[25] G.-M. Zhang, Y.-F. Yang, and F.-C. Zhang, arXiv:1909.11845 (2019).

[26] N. Singh, arXiv:1909.07688 (2019).

[27] A. V. Krukau, O. A. Vydrov, A. F. Izmaylov, and G. E. Scuseria, J. Chem. Phys. **125**, 224106 (2006).

[28] B. G. Janesko, T. M. Henderson, and G. E. Scuseria, Phys. Chem. Chem. Phys. **11**, 443 (2009).

[29] J. P. Perdew, K. Burke, and M. Ernzerhof, Phys. Rev. Lett. **77**, 3865 (1996).

[30] G. Kresse and J. Furthmüller, Comput. Mater. Sci. **6**, 15 (1996).

[31] G. Kresse and J. Furthmüller, Phys. Rev. B **6**, 15 (1996).

[32] P. Blaha, K. Schwarz, G. K. H. Madsen, D. Kvasnicka, J. Luitz, R. Laskowski, F. Tran, and L. D. Marks, *WIEN2k*: An Augmented Plane Wave and Local Orbitals Program for Calculating Crystal Properties (Vienna University of Technology, Austria, 2001).

[33] J. Kuneš, R. Arita, P. Wissgott, A. Toschi, H. Ikeda, and K. Held, Comput. Phys. Commun. **181**, 1888 (2010).

[34] V. J. Emery, Phys. Rev. Lett. **58**, 2794 (1987).

[35] V. J. Emery and G. Reiter, Phys. Rev. B **38**, 4547 (1988).

[36] See Supplemental Material for electronic band structures of LaNiO$_2$ under different schemes and effects, Wannier projection, and symmetry analysis.

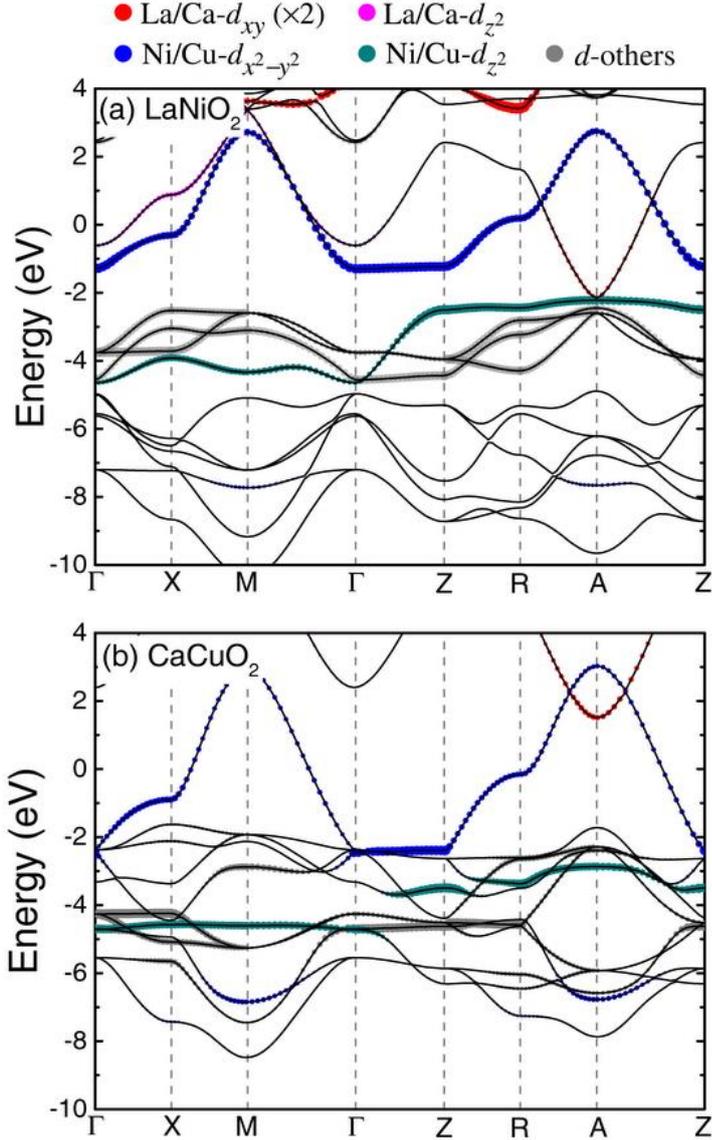

Fig. 1 Orbital projected electronic band structures of LaNiO$_2$ and CaCuO$_2$. The size of red, magenta, blue, dark cyan and grey dots represent the weight of La/Ca-$d_{xy}$, La/Ca-$d_{z^2}$, Ni/Cu-$d_{x^2-y^2}$, Ni/Cu-$d_{z^2}$, and other $d$ orbitals, respectively. The weight of La/Ca-$d_{xy}$ has been enlarged twice. The six bands within the energy range of −10 ~ −5 eV in LaNiO$_2$ without label of orbital projection are contributed by O-$p$ orbitals, while the corresponding bands in CaCuO$_2$ within the energy range of −9 ~ −4 eV are contributed by both O-$p$ and Ni-$d$ orbitals. The HSE functional was used in the calculation.

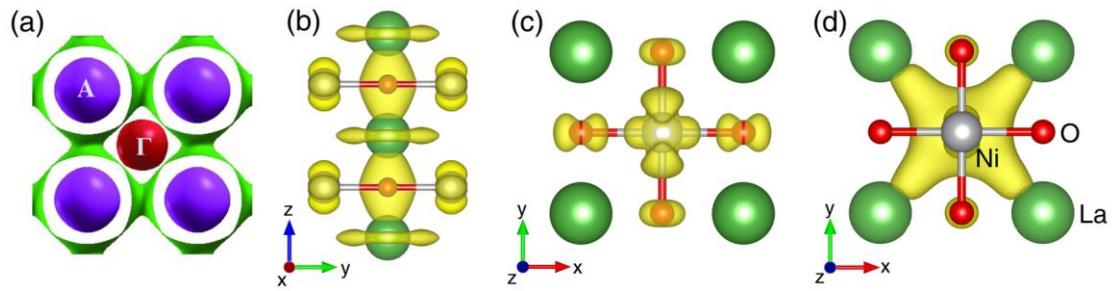

Fig. 2 (a) Fermi surface of LaNiO$_2$. Three different electron pockets around Γ, Z and A points are shown by red, purple and green spheres, respectively. (b)-(d) The charge density of these three electron pockets near Fermi level.

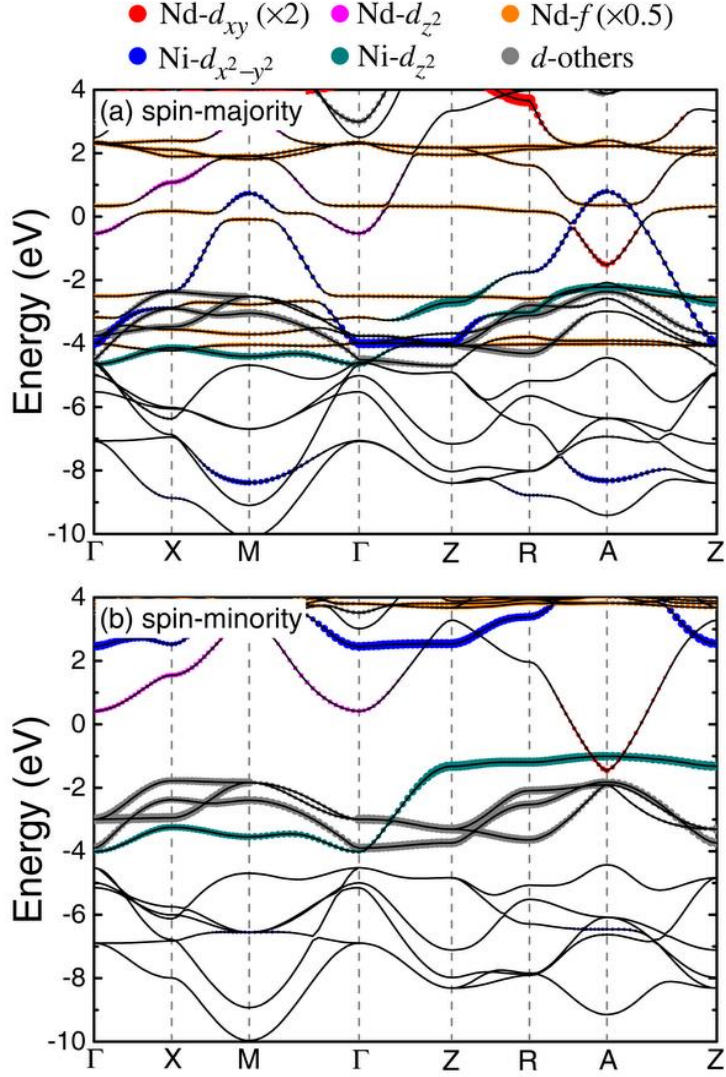

Fig. 3 Orbital projected electronic band structure of NdNiO$_2$. The size of red, magenta, blue, orange, dark cyan and grey dots represent the weight of Nd-$d_{xy}$, Nd-$d_{z^2}$, Nd-$f$, Ni-$d_{x^2-y^2}$, Ni-$d_{z^2}$ and other $d$ orbitals, respectively. The weight of Nd-$d_{xy}$ and Nd-$f$ orbitals have been enlarged twice and half time, respectively. The HSE functional and spin-polarization were considered in the calculation.

# Supplemental Material for "Electronic structure of rare-earth infinite-layer ReNiO$_2$ (Re=La, Nd)"

Peiheng Jiang[1,†], Liang Si[1,2†], Zhaoliang Liao[3], Zhicheng Zhong[1,4*]

[1] *Key Laboratory of Magnetic Materials and Devices & Zhejiang Province Key Laboratory of Magnetic Materials and Application Technology, Ningbo Institute of Materials Technology and Engineering, Chinese Academy of Sciences, Ningbo 315201, P. R. China*

[2] *Institut für Festkörperphysik, TU Wien, Wiedner Hauptstraße 8-10, 1040 Vienna, Austria*

[3] *National Synchrotron Radiation Laboratory, University of Science and Technology of China, Hefei 230026, P. R. China*

[4] *China Center of Materials Science and Optoelectronics Engineering, University of Chinese Academy of Sciences, Beijing 100049, P. R. China*

*Corresponding authors: Z.Z. (email: zhong@nimte.ac.cn)

† These authors contributed equally to this work.

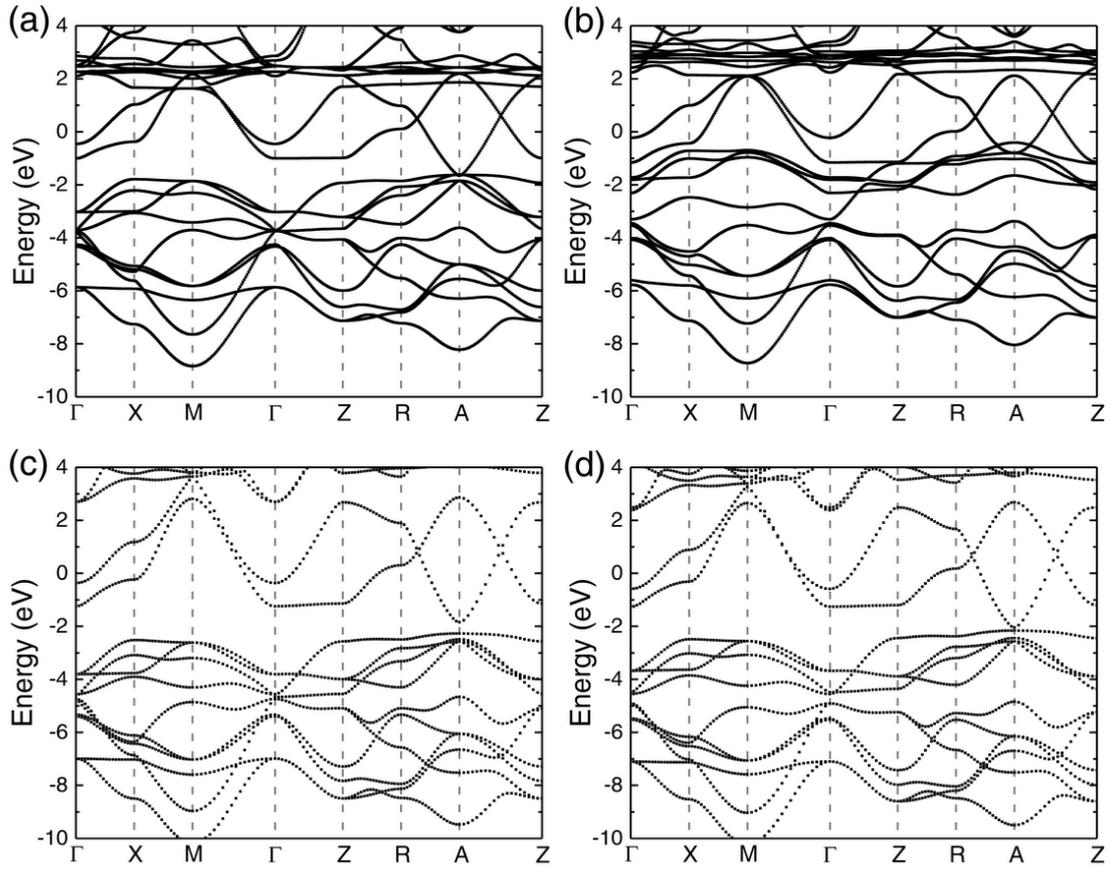

Fig. S1: Calculated electronic band structure of LaNiO$_2$. (a) DFT+U calculation without spin-orbit coupling. Here, U=2.0 and 5.0 eV are applied for La-5$d$ and Ni-3$d$, respectively. (b) DFT calculation with spin-orbit coupling. HSE calculation with (c) 0.2 hole/Ni doping; and (d) 0.6% tensile strain to simulate the substrate effect.

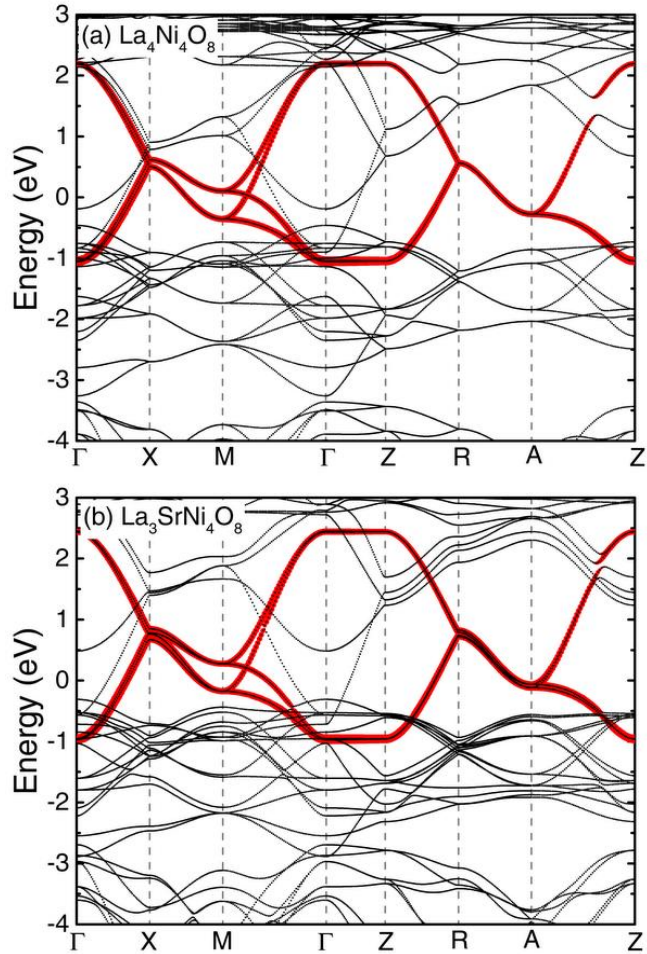

Fig. S2: Electronic band structures of LaNiO$_2$ with and without Sr doping. The size of red dots represent the weight of Ni-$d_{x^2-y^2}$ orbital. The Sr doping has negligible influence on Ni-$d_{x^2-y^2}$ orbital, and La-$d$ orbitals across Fermi level in both doped and un-doped systems.

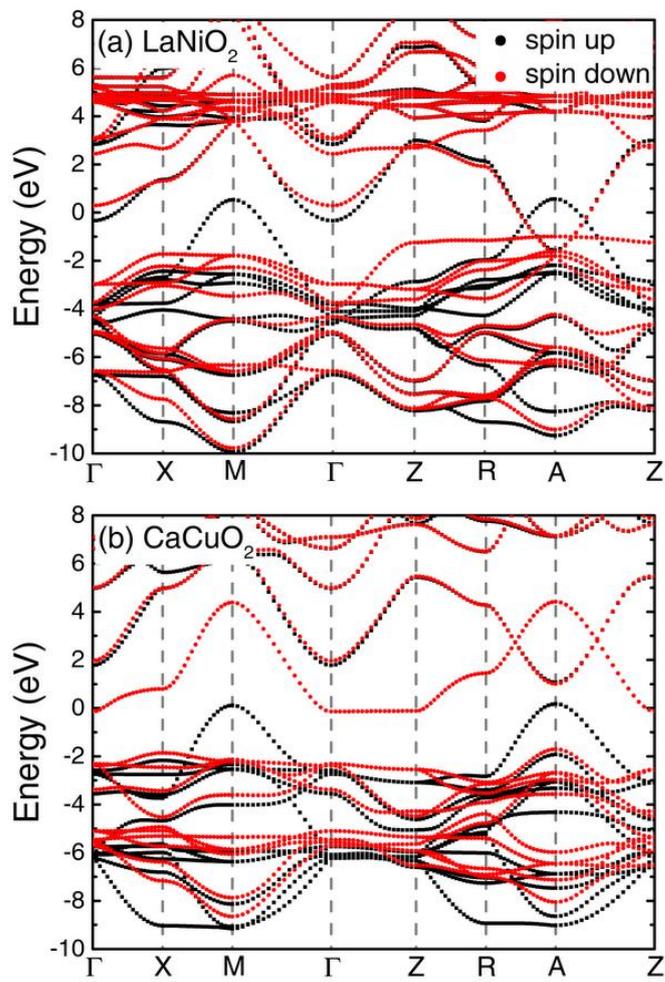

Fig. S3: Calculated spin polarized electronic band structure of LaNiO$_2$ and CaCuO$_2$. The HSE functional was used in calculations.

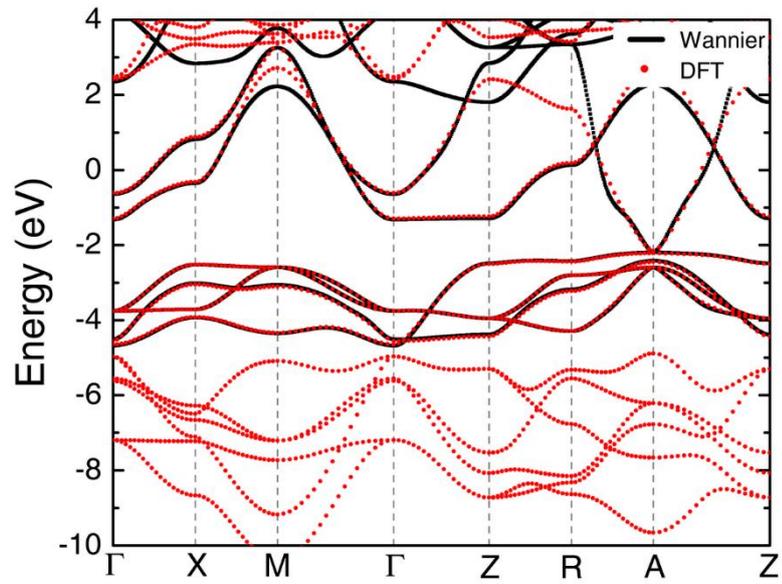

Fig S4: Electronic band structures of LaNiO$_2$ calculated by Wannier projection and density functional theory. The HSE functional was used in calculations.

Tabel I On-site hopping terms $t_{\alpha\beta}$ between each $d$ orbitals in LaNiO$_2$ calculated by Wannier projection with HSE functional.

| $t_{\alpha\beta}$ (eV) | La-$d_{xy}$ | La-$d_{xz}$ | La-$d_{yz}$ | La-$d_{x^2-y^2}$ | La-$d_{z^2}$ | Ni-$d_{xy}$ | Ni-$d_{xz}$ | Ni-$d_{yz}$ | Ni-$d_{x^2-y^2}$ | Ni-$d_{z^2}$ |
|---|---|---|---|---|---|---|---|---|---|---|
| La-$d_{xy}$ | 2.139 | 0 | 0 | 0 | 0 | -0.111 | -0.252 | -0.252 | 0 | -0.385 |
| La-$d_{xz}$ | | 4.296 | 0 | 0 | 0 | 0.131 | -0.104 | 0.139 | 0.052 | -0.088 |
| La-$d_{yz}$ | | | 4.296 | 0 | 0 | 0.131 | 0.139 | -0.104 | -0.052 | -0.088 |
| La-$d_{x^2-y^2}$ | | | | 4.146 | 0 | 0 | -0.181 | 0.181 | -0.021 | 0 |
| La-$d_{z^2}$ | | | | | 2.649 | 0.152 | 0.016 | 0.016 | 0 | 0.253 |
| Ni-$d_{xy}$ | | | | | | -3.167 | 0 | 0 | 0 | 0 |
| Ni-$d_{xz}$ | | | | | | | -3.077 | 0 | 0 | 0 |
| Ni-$d_{yz}$ | | | | | | | | -3.077 | 0 | 0 |
| Ni-$d_{x^2-y^2}$ | | | | | | | | | 0.437 | 0 |
| Ni-$d_{z^2}$ | | | | | | | | | | -2.938 |

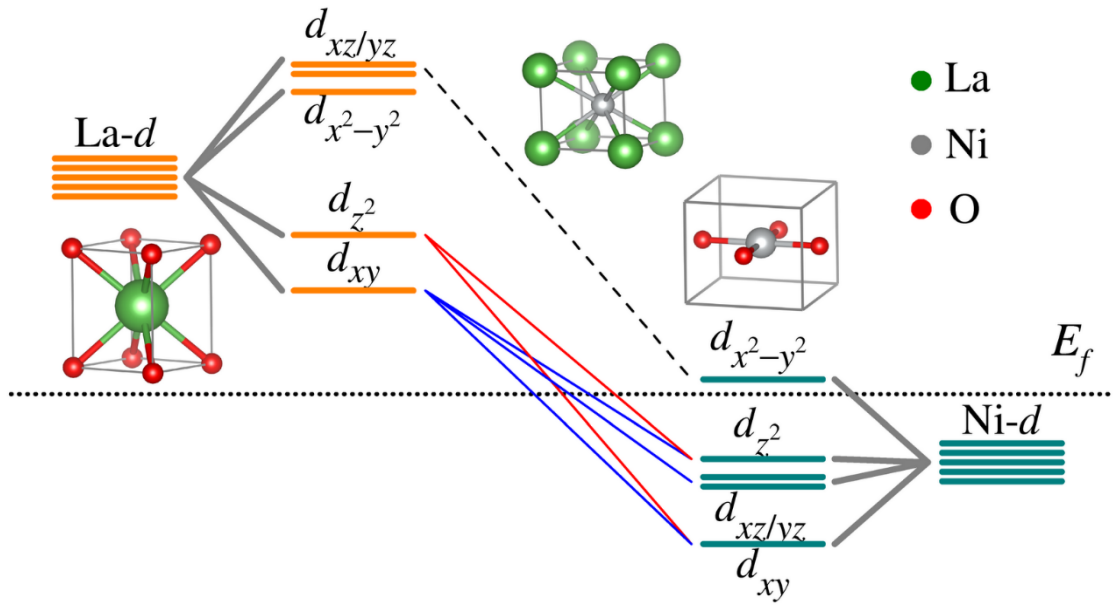

Fig S5: Crystal field splitting of LaNiO$_2$. The red and blue lines indicate large hopping terms while the grey dash line indicates a small hopping term between two orbitals.

Table II Hopping terms $t(\mathbf{R})$ of Ni-$d_{x^2-y^2}$ orbital in LaNiO$_2$ calculated by Wannier projection with HSE functional. $\mathbf{R}$ denotes the inter-atomic vector in the unit of lattice constant.

| $\mathbf{R}$ | (0,0,0) | (0,0,1) | (0,1,0) | (0,1,1) | (1,1,0) | (1,1,1) | (0,0,2) | (0,2,0) |
|---|---|---|---|---|---|---|---|---|
| $t(\mathbf{R})$ (eV) | 0.482 | -0.055 | -0.482 | -0.001 | 0.110 | 0.014 | 0.001 | -0.051 |

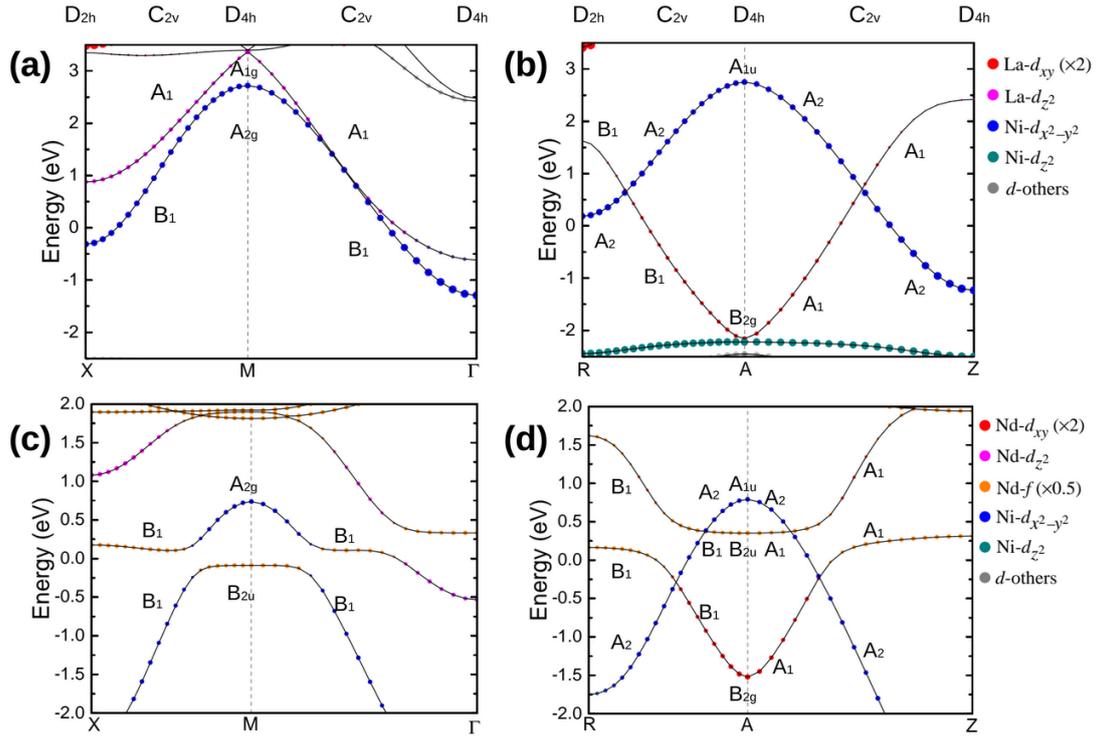

Fig S6: Band structures of LaNiO$_2$ (a-b) and NdNiO$_2$ (c-d) with different irreducible representations of along X-M-Γ (a, c) and R-A-Z (b, d) paths. The labels above the curves denote the wave vector groups at different k-points. For NdNiO$_2$ only the spin-majority channel is plotted for simplification.

Table III (a): Wave vector symmetry along Γ-plane ($k_z = 0$) of LaNiO$_2$. The k-path definition is same as in the bands plotting of Fig. 1 and Fig. 3 of main text. The No. 20 and No. 21 (the label numbers 20 and 21 originate from out Wien2k code computational outputs for LaNiO$_2$) denote the lower and upper bands at X point shown in Fig. S6 (a), respectively. The No. 20 band always label the band majorly contributed by Ni-$d_{x^2-y^2}$. The wave vector groups at different k-paths/points are listed. Mulliken labels of each irreducible representation are used. In the rows of "character", the major bands' contribution is listed. The full wave vector group analysis of NdNiO$_2$ is not shown here due to its complex bands nature and large amount of bands near Fermi energy.

| k-path/point | Γ | Γ-X | X | X-M | M | M-Γ | Γ |
|---|---|---|---|---|---|---|---|
| P-Group | $D_{4h}$ | $C_{2v}$ | $D_{2h}$ | $C_{2v}$ | $D_{4h}$ | $C_{2v}$ | $D_{4h}$ |
| No. 21 | $A_{1g}$ | $A_1$ | $A_{1g}$ | $A_1$ | $A_{1g}$ | $A_1$ | $A_{1g}$ |
| character | La-$d_{z^2}$ Ni-$d_{z^2}$ | La-$d_{z^2}$ | La-$d_{z^2}$ | La-$d_{z^2}$ | La-$d_{z^2}$ | La-$d_{z^2}$ Ni-$d_{z^2}$ | La-$d_{z^2}$ Ni-$d_{z^2}$ |
| No. 20 | $B_{1g}$ | $A_1$ | $B_{2u}$ | $B_1$ | $A_{2g}$ | $B_1$ | $B_{1g}$ |
| character | Ni-$d_{x^2-y^2}$ | Ni-$d_{x^2-y^2}$ | Ni-$d_{x^2-y^2}$ | Ni-$d_{x^2-y^2}$ | Ni-$d_{x^2-y^2}$ | Ni-$d_{x^2-y^2}$ | Ni-$d_{x^2-y^2}$ |

Table III (b): Wave vector symmetry along Z-plane ($k_z = 0.5$).

| k-path/point | Γ-Z | Z | Z-R | R | R-A | A | A-Z | Z |
|---|---|---|---|---|---|---|---|---|
| P-Group | $C_{4v}$ | $D_{4h}$ | $C_{2v}$ | $D_{2h}$ | $C_{2v}$ | $D_{4h}$ | $C_{2v}$ | $D_{4h}$ |
| No. 21 | $A_1$ | $A_{1g}$ | $A_1$ | $B_{2u}$ | $B_1$ | $B_{2g}$ | $A_1$ | $A_{1g}$ |
| character | La-$d_{z^2}$ Ni-$d_{z^2}$ | La-$f$ | La-$f$ | La-$f$ | La-$d_{xy}$ | La-$d_{xy}$ | La-$d_{xy}$ | La-$f$ |
| No. 20 | $B_1$ | $B_{2u}$ | $B_1$ | $B_{3g}$ | $A_2$ | $A_{1u}$ | $A_2$ | $B_{2u}$ |
| character | Ni-$d_{x^2-y^2}$ | Ni-$d_{x^2-y^2}$ | Ni-$d_{x^2-y^2}$ | Ni-$d_{x^2-y^2}$ | Ni-$d_{x^2-y^2}$ | Ni-$d_{x^2-y^2}$ | Ni-$d_{x^2-y^2}$ | Ni-$d_{x^2-y^2}$ |